# Grant-Free Power Allocation for Ultra-Dense Internet of Things Environments: A Mean-Field Perspective⋆

Sami Nadif[a], Essaid Sabir[b,∗], Halima Elbiaze[b] and Abdelkrim Haqiq[a]

[a]Computer Networks, Mobility and Modeling Laboratory (IR2M), FST, Hassan I University of Settat, 26000, Morocco
[b]Department of Computer Science, University of Quebec at Montreal (UQAM), Montreal, H2L 2C4, Quebec, Canada

ARTICLE INFO

*Keywords*:
Internet of Things (IoT)
Grant-Free Access
Massive IoT
Ultra-Dense Networks
Power Allocation
Mean-Field Game

ABSTRACT

Grant-free access, in which each Internet-of-Things (IoT) device delivers its packets through a randomly selected resource without spending time on handshaking procedures, is a promising solution for supporting the massive connectivity required for IoT systems. In this paper, we explore grant-free access with multi-packet reception capabilities, with an emphasis on ultra-low-end IoT applications with small data sizes, sporadic activity, and energy usage constraints. We propose a power allocation scheme that integrates the IoT device's traffic and energy budget by using a stochastic geometry framework and mean-field game theory to model and analyze mutual interference among active IoT devices. We also derive a Markov chain model to capture and track the IoT device's queue length and derive the successful transmission probability at steady-state. Simulation results illustrate the optimal power allocation strategy and show the effectiveness of the proposed approach.

## 1. Introduction

Massive Internet of Things (IoT) communications are essential in the fifth-generation (5G) and beyond cellular networks [1]. They are distinguished by a large number of low-cost IoT devices that operate mostly in the uplink and have small packets, sporadic activity, and restrictive energy usage requirements. Hence, massive IoT requires a whole different set of Medium Access Control (MAC) protocols than those intended for human-centric communications. To be more specific, existing cellular networks use Grant-Based (GB) transmission, which means that each IoT device must conduct a handshake procedure to establish a connection with the Base Station (BS) anytime a new packet needs to be sent. Furthermore, the handshake procedure involves exchanging multiple signaling messages (i.e., scheduling request, scheduling grant, and resource allocation) to facilitate exclusive channel access, which might take several tens of milliseconds citec1. However, as packet sizes shrink and the number of IoT devices grows, this handshake procedure becomes inefficient, potentially resulting in overhead signaling and radio access congestion [2]. Additionally, the number of signaling messages has a significant impact on the energy efficiency of IoT devices (shorter transmissions preserve energy). To address these issues, Grant-Free (GF) transmission [3, 4, 5, 6] is a promising solution. This approach removes the need for a handshaking process by allowing each IoT device to send its packets across a randomly selected resource, without prior coordination with the BS. SigFox and LoRa are two examples of low-power wide-area network technologies that implement GF transmission for efficient IoT connectivity [7, 8]. This is made possible through the use of low-complexity contention-based random access schemes, such as variations of the well-known ALOHA protocol [9, 10], making GF transmission ideal for large-scale IoT networks with infrequent communication needs. However, since the orthogonal resources are not allocated by the base station, numerous IoT devices may use the same resource for transmission, potentially resulting in a collision. To reduce the effects of collisions, non-orthogonal multiple access protocols can be used in conjunction with random access schemes [3, 11]. More specifically, the Multi-Packet Reception (MPR) capability [12, 13] given by a non-orthogonal transmission mechanism enables numerous IoT devices to use the same resource and transmit their packets concurrently, reducing the occurrence of packet collisions. Nevertheless, the performance of grant-free

∗Corresponding author
s.nadif@uhp.ac.ma (S. Nadif); sabir.essaid@uqam.ca (E. Sabir); elbiaze.halima@uqam.ca (H. Elbiaze); abdelkrim.haqiq@uhp.ac.ma (A. Haqiq)
ORCID(s): 0000-0003-4141-5491 (S. Nadif); 0000-0001-9946-5761 (E. Sabir); 0000-0001-5681-6445 (H. Elbiaze); 0000-0002-8857-6586 (A. Haqiq)

access with MPR capability degrades in a massive IoT environment with sporadic traffic [14]. Thus, in order to take full advantage of this solution, a distributed resource management technique such as power control [15, 16, 17] must be considered. Power control was originally employed in cellular networks to manage interference, but it's also a flexible mechanism to provide quality of service and reduce energy consumption. Thus, grant-free transmission, when combined with an effective power allocation strategy, can therefore meet the 5G standards for IoT devices, which include a battery life of more than ten years [18].

In this paper, we consider a large-scale ultra-dense IoT networks using grant-free transmission with $J$-MPR capability (assuming that up to $J$ collided IoT devices can be decoded on a single resource), and we focus on ultra-low-end IoT applications with small data sizes, sporadic activity, and restrictive energy usage requirements. We develop an analytical model that takes into account spatial randomness and temporal traffic generation. For tractability of our analysis, both the BSs and the IoT devices are modeled using a homogeneous Poisson Point Process (PPP) where the BSs boundaries can be shown by a weighted Voronoi tessellation. From the temporal perspective, we consider a sporadic IoT traffic where packet arrival at each IoT device is governed by a Bernoulli process with small arrival rate. The main difficulty in considering a spatiotemporal model is that the set of active IoT devices that cause interference changes dramatically over time. Thus, to assess the uplink performance of the grant-free transmission in a massive IoT network, manage inter-cell interference, and avoid unnecessary energy wastage, we propose a distributed uplink power allocation strategy under spatiotemporal fluctuation. The power allocation problem is initially treated as a differential game due to the coupling in interference. Then, by using stochastic geometry framework and Mean-Field Game (MFG) theory to model and analyze mutual interference among active IoT devices, we extend the differential game to a MFG. The MFG framework enables each IoT device to determine its optimal power allocation strategy based on its own energy budget and the statistical distribution of the system state, known as the mean-field. Moreover, we develop a Markov chain model to monitor the IoT device's queue length and derive the steady-state successful transmission probability. Finally, by formulating the problem as a mean-field optimal control problem, we can obtain a set of equations that allow us to achieve the mean-field equilibrium through iterative solution.

### 1.1. Related Work

The use of game theory in the analysis of random access schemes has been widely studied over the past decades [19, 20, 21, 22]. Game theory provides a way to analyze the decision-making process between multiple individuals or entities who are interacting with each other, and it has been used to study various aspects of random access schemes, such as the users behavior, and system performance optimization. The authors in [19], for example, use game theory to analyze the Aloha protocol from the perspective of selfish users who have two possible actions: transmit or wait. The authors construct an Aloha game to study the optimal behavior of individual users and show that the Aloha game has an equilibrium. The authors in [20] analyze the ALOHA protocol with users having two transmission power levels and use two non-cooperative optimization concepts, the Nash equilibrium and the evolutionary stable strategy. The performance of these concepts is compared with a cooperative solution and the impact of multiple power levels is analyzed. Game theory has also been used to capture the interactions between a set of IoT devices and predict their optimal strategies on contemporary IoT networks by analyzing the Nash and/or Stackelberg equilibrium [23]. The authors in [24] use the Nash equilibrium to provide an energy-efficient access point allocation in an IoT network. In [25], the authors use the age of information metric to analyze the competition between multiple IoT devices, who can choose their own transmission probability, in an irregular repetition slotted Aloha IoT system. By analyzing the Nash equilibrium, they highlight that introducing a transmission cost can regulate the overall performance. In contrast, related works based on the Stackelberg equilibrium concept include [26, 27], which develop and analyze hierarchical game models for IoT networks based on the leader/follower principle. The authors in [26], for example, develop a multi-leader/multi-follower Stackelberg game to provide a caching strategy in a 5G-enabled IoT network by analyzing a competitive scenario with various 5G mobile network operators and different content providers.
However, when dealing with a large number of IoT devices, the atomic equilibria concepts become extremely challenging. In this circumstance, it is more interesting for an IoT device to deal with collective behavior than with the specific individual strategy of each IoT device. Hence, the Mean-Field Game (MFG) theory [28, 29, 30, 31, 32] has increasingly gained attention in massive IoT networks. Related works analyzing the mean-field equilibrium in massive IoT networks could be found in [33, 34, 35, 36]. For example, the authors in [34] investigate delay-optimal random access in large-scale energy harvesting IoT networks. They handle the coupling between the data and energy queues using a two-dimensional Markov decision process, and they employ the MFG theory to disclose the coupling among IoT devices by exploiting the large-scale property. In [36], under the grant-free communication framework, the age of information minimization problem is analyzed in a massive IoT network using a mean-field evolutionary game-based approach by optimizing packet sampling and scheduling.

The works related to grant-free transmission, on the other hand, include [5], where a Semi-Granted Multiple Access (SGMA) approach is analyzed for non-orthogonal multiple access in 5G networks, allowing grant-based and grant-free transmission to share the same wireless resource. Then, a heuristic SGMA resource allocation algorithm is presented to enhance network capacity and user connectivity. The authors in [37] present a grant-free user activity detection scheme in a massive IoT network with extremely low complexity and latency. They use multiple antennas at the base station to generate spatial filtering via a fixed beamforming network, which reduces inter-beam interference. They also suggest using orthogonal multiple access technology to minimize intra-beam interference in the temporal domain. In [4], the authors investigate an asynchronous grant-free transmission protocol with the goal of reducing energy consumption and delay by relaxing the synchronization requirement at the cost of sending multiple packet replicas and using a more complex signal processing technique. The suggested approach is scrutinized by developing closed-form expressions for critical performance characteristics, including reliability and battery life. Related works that investigate a spatiotemporal Modeling [38, 39] for grant-free transmission can be found in [40, 41]. The authors in [40] analyze three grant-free transmission schemes that use Hybrid Automatic Repeat reQuest (HARQ) retransmissions: Reactive, K-repetition, and Proactive. They provide a spatiotemporal assessment model for contention-based grant-free transmission and define the access failure probability to evaluate the reliability and latency performance under the three grant-free HARQ schemes. In [41], the authors investigate the energy efficiency and packet transmission probability in grant-free uplink IoT networks. To do this, a spatiotemporal model is constructed leveraging queueing theory and a stochastic geometry framework, where each device is represented by a two-dimensional discrete-time Markov chain and expected to use full path loss inversion power control with a target power level. In a similar vein, our work also examines grant-free transmission using a spatiotemporal model. However, instead of relying on path loss inversion power control, we employ a power allocation strategy based on mean-field game theory. Within this framework, we also aim to analyze energy efficiency and packet transmission probability.

### 1.2. Our Contributions

The contributions of this paper are manifold and can be summarized as follows:

(1) We present a spatiotemporal model for grant-free transmission with $J$-MPR capability by using stochastic geometry and Markov chain theory. From the spatial perspective, stochastic geometry is applied to model and analyze the mutual interference among active IoT devices (i.e., those with at least one packet in their queue). From the temporal perspective, Markov chain theory is used to model the correlation of the number of packets in a queue over different frames.

(2) To address the issue of the large number of IoT devices, we present a mean-field power allocation scheme that integrates the IoT device's traffic generation and energy budget. Our approach enables IoT devices to distributively compute their power allocation control strategies without having complete awareness of the strategies or states of other IoT devices.

(3) We combine stochastic geometry analysis and mean field formulation to derive the successful transmission probability in a massive grant-free IoT network (see proposition 2). The success probability in the paper's model is influenced by multiple factors, such as IoT density, arrival rate, base station density, and multi-packet reception capability.

(4) We assess the performance of a massive grant-free IoT network with MPR capability in terms of packet transmission success probability and average delay.

The rest of the paper is organized as follows: The section 2 presents the system model and the assumptions. The power allocation problem is formalized in section 3. The section 4 introduce the discrete-time

Markov chain model. The section 5 provides the numerical and simulation results. Finally, concluding remarks are given in section 6.

## 2. System Model

We consider a massive IoT network using an orthogonal multiple access scheme, where the total bandwidth is divided into $L$ orthogonal channels, denoted by $\mathbf{L} = \{\ell_i, i = 1, \dots, L\}$. We presume that the frequency resource is reused throughout the network (frequency reuse factor of 1), which generates intercell interference. Furthermore, in this paper, we consider a single-tier of base stations (BS) with Multi-Packet Reception (MPR) capability that are spatially distributed according to a homogeneous PPP denoted by $\phi_s = \{\mathbf{s}_i, i = 1, 2, \dots\}$ with intensity $\lambda_s$, where $\mathbf{s}_i$ is the location of the $i^{th}$ BS. We consider a $J$-MPR model where up to $J$ IoT devices can be decoded on a single channel. We assume that if $j$ IoT devices choose the same channel, there is no packet collision whenever $j \leq J$, whereas all the $j$ IoT devices are not decoded (collided) whenever $j > J$. The IoT devices, on the other hand, are randomly distributed and modeled by a homogeneous PPP denoted by $\phi_u = \{\mathbf{u}_i, i = 1, 2, \dots\}$ with intensity $\lambda_u$, where $\mathbf{u}_i$ is the location of the $i^{th}$ IoT device. Each IoT device is served via its geographically nearest BS. Thus, the BS boundaries can be shown by a weighted Voronoi tessellation. An important random parameter is the distance $r$ separating an IoT device from its serving BS. Since each IoT device communicates with the closest BS, no other base station can be closer than $r$. Thus, the distance of an arbitrary IoT device from its serving BS has a cumulative distribution function given as follows:

$$\begin{aligned} F_r(r_0) &= \mathbb{P}[r \leq r_0] \\ &= 1 - \mathbb{P}[\text{No BS closer than } r_0] \\ &= 1 - e^{-\lambda_s \pi r_0^2}. \end{aligned} \tag{1}$$

Thus, the probability density function can be found as

$$f_r(r) = \frac{dF_r(r)}{dr} = 2\pi\lambda_s r e^{-\lambda_s \pi r^2}. \tag{2}$$

Let $V$ be the area of a Voronoi cell. The number of IoT devices associated with a specific BS of area $V = v$, defined as $N_v$, follows a Poisson distribution with the probability mass function given by:

$$\mathbb{P}[N_v = k | V = v] = \frac{(\lambda_u v)^k}{k!} e^{-\lambda_u v}, \quad k = 0, 1, \dots \tag{3}$$

Moreover, the Voronoi cell area $V$ is a random variable that can be approximated by the gamma distribution with shape $c = 3.575$ and rate $\lambda_s c$ [42]. The corresponding probability density function is:

$$f_V(v) = \frac{v^{c-1}(\lambda_s c)^c}{\Gamma(c)} e^{-(\lambda_s c)v}, \quad v > 0. \tag{4}$$

From the temporal perspective, we assume that the network operates in a synchronized manner and that the timeline is segmented into frames with duration $T_f$. This simplifies the user detection procedure since multiple packets are allowed to be sent on the same channel simultaneously. We also assume that the IoT devices use grant-free transmission. In other words, the packet for each IoT device will be transmitted immediately once it is generated. In this paper, the packet arrival process at each IoT device is modeled by a Bernoulli process, which is commonly utilized in discrete-time system modeling, with a small arrival rate $0 \leq p_a \leq 1$ (sporadic activity). Note that $p_a$ is also the probability that an IoT device will generate a packet in a particular frame. Furthermore, IoT devices with a non-empty queue may employ a frame for a single packet uplink transmission attempt. As a result, in each frame, only one packet may arrive and/or depart from the queue of each IoT device in the network. Each IoT device transmits packets via a first-come, first-served packet scheduling scheme and has a queue that can store a maximum of $M$ packets. We also assume that the generated packet is sufficiently small and thus can be successfully transmitted through each transmission attempt if there is no packet collision and the received Signal to Interference-plus-Noise Ratio (SINR) is greater than a threshold $\theta$. Furthermore, we presume that BS employs access barring to control overall traffic load in the system, where IoT devices with at least one packet in their queue, referred to as active IoT devices, attempt uplink transmissions in the current frame with probability $1 - p_b$ or skip it with probability $p_b$.

In this work, we denote by $\pi_a$ the probability that an IoT device is active, for which we will derive a closed-form expression in section 4. The active IoT devices that attempts an uplink transmissions in a given frame using channel $\ell_i \in \mathbf{L}$ constitute a PPP, denoted by $\phi_a = \{\mathbf{a}_i, i = 1, 2, \dots\}$, with intensity $\lambda_a = ((1-p_b)\pi_a/L)\lambda_u$. Let's $N_a$ be the number of active IoT devices in a given Voronoi cell using channel $\ell_i \in \mathbf{L}$. Therefore, by using (3) and (4), the unconditional probability mass function of $N_a$, for $k = 0, 1, 2, \dots$, writes

$$\begin{aligned}\mathbb{P}[N_a = k] &= \int_0^\infty \frac{(\lambda_a v)^k}{k!} e^{-\lambda_a v} f_V(v)\, dv \\ &= \frac{\Gamma(k+c)}{\Gamma(k+1)\Gamma(c)} \cdot \frac{(\lambda_a)^k (\lambda_s c)^c}{(\lambda_a + c\lambda_s)^{k+c}}.\end{aligned} \quad (5)$$

In the rest of this paper, we consider a large circle of radius $R$ to be the spatial domain of our analysis, denoted as $\mathcal{C}_R$. It is worth noting that the number of active IoT devices attempting an uplink transmission in a given frame using channel $\ell_i$ in $\mathcal{C}_R$, denoted as $\mathcal{N}_R$, is a Poisson random variable with a mean intensity $\lambda_a \pi R^2$.

The outcome of a transmission is assessed through the received time-varying SINR. Without loss of generality, the experienced SINR under a Gaussian single input, single output channel writes:

$$\Gamma_i(t, P_i, \mathbf{P}_{-i}) = \frac{P_i(t) H_{i,i}(t) D_{i,i}(r)}{\sigma_0 + I_i(t, \mathbf{P}_{-i})}, \quad (6)$$

where in the above expression, $P_i \in [0, P_{max}]$ is the transmit power of IoT device $i$, $\mathbf{P}_{-i}$ denotes the transmit power vector of the active IoT devices using the same channel without $i$, $H_{i,j}$ (resp. $D_{i,j}$) is a parameter representing the multipath fading (resp. path-loss ) between the IoT device $j$ and BS serving the IoT device $i$, $\sigma_0$ is the noise power, and $I_i(t, \mathbf{P}_{-i})$ denotes the interference caused by the active IoT devices using the same channel for transmission expressed as:

$$I_i(t, \mathbf{P}_{-i}) = \sum_{j=1, j\neq i}^{|\mathcal{N}_R|} P_j(t) H_{i,j}(t) D_{i,j}(r). \quad (7)$$

Finally, we assume that the channel between all the transmitters and all the receivers experiences an independent Rayleigh fading $H$, exponentially distributed with unity mean, and a path-loss $D(r) = r^{-\alpha}$ with exponent $\alpha > 2$.

## 3. Power Allocation: A Mean-Field Approach

### 3.1. Differential Game Model

The differential power allocation game is played over time $t \in [0, T_f]$, where $T_f$ represents the frame duration.

• **Player sets:** $\mathcal{N}_R = \{1, 2, \dots, |\mathcal{N}_R|\}$, the active IoT devices that attempts an uplink transmissions in a given frame using channel $\ell_i \in \mathbf{L}$

• **State:** The state of an IoT device $i$ time $t$ is described by its remaining energy at that time, given by $E_i(t) \in [0, E_{i,max}]$, evolving according to the following differential equation:

$$dE_i(t) = -P_i(t)\, dt, \quad (8)$$

where $P_i(t)$ is the transmit power, and $E_{i,max} = E_i(0)$ is the energy budget fixed by the IoT device $i$ to spend over $[0, T_f]$. The dynamics (8) implies that the variation in the energy budget during $dt$ is proportional to the transmission power.

• **Actions:** Transmit power $P_i(t) \in [0, P_{max}]$ which is allowed to depend not only on time, but on its own state $E_i(t)$ and on the states of all other active IoT devices in the system at time $t$, denoted as $\mathbf{E}_{-i}(t)$. A power allocation strategy of the IoT device $i$ will be denoted by $P_i$ with $P_i(t) = P_i(t, E_i(t), \mathbf{E}_{-i}(t))$.

• **Utility function:** The goal of each IoT device is to adapt its actions according to its remaining energy while maximizing its throughput. Thus, the average utility function of the IoT device $i$ is given by:

$$U_i(P_i, \mathbf{P}_{-i}, p_s) = \mathbb{E}\left[\int_0^{T_f} F_i(t, P_i, \mathbf{P}_{-i}, p_s)\, dt\right], \quad (9)$$

where

$$F_i := (1 - p_s) P_i - p_s \log_2(1 + \Gamma_i), \quad (10)$$

and $p_s$ is the successful transmission probability (see proposition 2).

Such utility function is especially relevant when the IoT devices have to trade-off between achieving the highest possible throughput and expending as little power as necessary.

• **Nash equilibrium:** A power allocation strategy profile $\mathbf{P}^* = (P_1^*, P_2^*, \dots, P_{|\mathcal{N}_R|}^*)$ is a feedback Nash equilibrium of the dynamic differential game if and only if $\forall i$, $P_i^*$ is a solution of the following optimal control problem:

$$\inf_{P_i} U_i(P_i, \mathbf{P}_{-i}^*, p_s), \quad (11)$$

subject to

$$d\begin{bmatrix} E_i(t) \\ \mathbf{E}_{-i}(t) \end{bmatrix} = \begin{bmatrix} -P_i(t) \\ -\mathbf{P}_{-i}^*(t) \end{bmatrix} dt, \quad (12)$$

To obtain the optimal power allocation strategies, the standard solution concept consists of analyzing the Nash equilibrium. However, the complexity of this approach increases with the number of IoT devices. Furthermore, it necessitates that each IoT device be fully aware of the states and actions of all other IoT devices, resulting in a tremendous volume of information flow. This is not feasible and impractical for a grant-free massive IoT network. Nevertheless, since the effect of other IoT devices on a single IoT device's average utility function is only via interference, it is intuitive that, as the number of IoT devices increases, a single IoT device has a negligible effect. Thus, we suggest using a mean-field limit for this game to convert these multiple interactions (interference) into a single aggregate interaction known as mean-field interference. However, the mean-field limit is only significant if the associated approximation error is small. It has been shown that, under appropriate conditions, the mean-field limit realizes an $\epsilon$-Nash equilibrium for the dynamic differential game, with $\epsilon$ converging to zero as the number of players goes to infinity [32, 43]. Therefore, in this paper, we consider a large scale ($R \to \infty$) massive IoT network under the assumption of frequency reuse factor of 1 to assure a small approximation error at the mean-field limit. It is important to note that if the frequency resources are not reused, increasing the number of BS in the network would decrease the number of IoT devices using the same channel for transmission, which negatively affect the accuracy of the mean-field limit.

### 3.2. Mean-Field Regime

The general setting of a mean-field regime is based on the following assumptions:

- The existence of large number of IoT devices ensured by considering large scale massive IoT network.
- Interchangeability: the permutation of the state (energy budget) among the IoT devices would not affect the optimal power allocation strategy. To guarantee this property, we assume that each IoT device only knows its individual energy budget and implements a homogeneous transmit power $P_i(t) = P(t, E_i(t))$.
- Finite mean-field interference $I_{mf}$ (see proposition 1).

Let $[0, E_{max}]$ be the energy domain of our analysis. We define the empirical energy distribution of the IoT devices in $C_R$ at time $t$ in $[0, T_f]$ as:

$$M(t,e) = \frac{1}{|\mathcal{N}_R|} \sum_{i=1}^{|\mathcal{N}_R|} \delta_e(E_i(t)), \quad \forall e \in [0, E_{max}], \tag{13}$$

where $\delta_e$ is the Dirac measure.

The basic idea behind the mean-field regime is to approximate a finite population with an infinite one, where the empirical energy distribution $M(t,e)$ almost surely converges, as $|\mathcal{N}_R| \to \infty$, to the probability density function $m(t,e)$ of a single IoT device, due to the strong law of large numbers. We will refer to the energy distribution $(m_t)_{t\geq 0}$ as the mean-field. Additionally, as the IoT devices become essentially indistinguishable, we can focus on a generic IoT device by dropping its index $i$ where its individual dynamic is written as:

$$\begin{cases} dE(t) = -P\big(t, E(t)\big)\, dt, \quad t \geq 0, \\ E(0) = E_0. \end{cases} \tag{14}$$

Thus, the evolution of the mean-field $(m_t)_{t\geq 0}$ over time $t$ in $[0, T_f]$ is described by a first-order partial differential equation, known as Fokker-Planck Kolmogorov (FPK) equation, given by [35]:

$$\begin{cases} \partial_t m(t,e) - \partial_e\big(P(t,e)m(t,e)\big) = 0, \\ m(0,.) = m_0. \end{cases} \tag{15}$$

The mean-field regime describes the mass behaviors of IoT devices in a massive IoT network, allowing the generic IoT device to determine its optimal power allocation strategy based only on its own energy budget and the initial mean-field. By expressing the interference in terms of an expectation over the mean-field that changes with time according to the FPK equation, we achieve a remarkable degree of economy in the description of population dynamics.

**Proposition 1.** *By following a stochastic geometry-based approach, the mean-field interference in large scale network for a generic IoT device is derived for $t \in [0, T_f]$ as follows:*

$$I_{mf}(t, \pi_a) = 2\pi\lambda_u \frac{(1-p_b)\pi_a}{L} \left[\frac{1}{2} + \frac{1}{\alpha - 2}\right] P_{mf}(t), \tag{16}$$

*where*

$$P_{mf}(t) = \int_0^{E_{max}} P(t,e)m(t,e)\, de. \tag{17}$$

Proof. The proof is given in Appendix A.1.

Now, we turn our attention to determining the SINR and the utility function, which are solely dependent on a generic IoT device's individual transmit power and the mean-field. The new parameters of the game are

defined as:

• Mean-field SINR: Since the distance of a generic IoT device from its serving BS has a probability density function given by (2) with mean $1/(2\sqrt{\lambda_s})$, we define the mean-field SINR as:

$$\Gamma_{mf}(P, I_{mf}) = \frac{P(t, E)(2\sqrt{\lambda_s})^{\alpha}}{\sigma_0 + I_{mf}(t, \pi_a)}. \tag{18}$$

• Mean-field utility function: The mean-field utility functions for a generic IoT device is generalized as follows:

$$U_{mf}(P, I_{mf}, p_s) = \int_0^{E_{max}} \int_0^{T_f} F_{mf}(P, I_{mf}, p_s) m(t, e) \, dt \, de, \tag{19}$$

where

$$F_{mf} := (1 - p_s)P - p_s \log_2(1 + \Gamma_{mf}). \tag{20}$$

### 3.3. Mean-Field Optimal Control

The mean-field optimal control problem of a generic IoT device is derived based on (11) and consists in finding the optimal power allocation strategy $P^*$ and the mean-field at the equilibrium $m^*$ satisfying:

$$\inf_P U_{mf}(P, I^*_{mf}, p_s) \tag{21}$$

where $I^*_{mf}$ is the mean-field interference at the equilibrium, by assuming that the interfering IoT devices use their optimal power allocation strategy and $m$ is a solution of

$$\begin{cases} \partial_t m(t, e) - \partial_e\big(P(t, e)m(t, e)\big) = 0, \\ m(0, .) = m_0. \end{cases} \tag{22}$$

Since $F$ is convex in $P$, the mean-field optimal control is a convex optimization problem. Therefore, the first-order optimality conditions are necessary and sufficient for the mean-field equilibrium.

#### *3.3.1. First-Order Optimality Conditions:*

The first-order optimality conditions of the mean-field optimal control problem are derived using the adjoint method. Note that even though this approach is used formally in the following, it can be made rigorous. We refer the interested reader to [44] for a rigorous derivation of these first-order optimality conditions.

Let's start by defining the Lagrangian of the minimization problem (21) under the constraint (22) as

$$\mathcal{L}(P, m, \mu) = U_{mf}(P, I^*_{mf}, p_s) \quad - \int_0^{E_{max}} \int_0^{T_f} \mu(t, e) \left(\partial_t m(t, e) - \partial_e\big(P(t, e)m(t, e)\big)\right) dt \, de, \tag{23}$$

where $\mu(t, e)$ represent the Lagrange multiplier.

The minimization problem (21) can be rewritten as a saddle-point problem:

$$\inf_{(P,m)} \sup_{\mu} \mathcal{L}(P, m, \mu). \tag{24}$$

By using the integration by parts, the first-order conditions characterizing the saddle-point $(P^*, m^*, \mu^*)$ of $\mathcal{L}$ are expressed as (22) together with:

$$\partial_P F_{mf} - \partial_e \mu^* = 0, \tag{25}$$

$$\partial_t \mu^* - P^* \partial_e \mu^* + F_{mf} = 0, \tag{26}$$

$$\mu^*(T_f, .) = 0.. \tag{27}$$

Note that the equation (26) reflects the adjoint equation of (22), popularly known as the Hamilton-Jacobi-Bellman equation in mean-field game theory.

Finally, the mean-field equilibrium can be obtained as the solution of the following mean-field system:

$$\begin{cases} \partial_t m - \partial_e(Pm) = 0, \quad m(0, .) = m_0, \\ \partial_t \mu^* - P^* \partial_e \mu^* + F_{mf} = 0, \quad \mu^*(T_f, .) = 0, \\ \partial_P F_{mf} - \partial_e \mu^* = 0, \end{cases} \tag{28}$$

which consists of two coupled partial differential equations, one evolving forward in time (the Fokker-Planck Kolmogorov) and the other one evolving backward in time (the Hamilton-Jacobi-Bellman equation).

#### *3.3.2. The Algorithm*

The mean-field system (28) is solved iteratively until the convergence point is reached to achieve the mean-field equilibrium. We employ a successive sweep method, which entails generating a series of nominal solutions $P_0, P_1, \dots, P_k, \dots$ that converges to the optimal power allocation strategy $P^*$. This iterative approach, which has proven to be effective in [45], is summarized in algorithm 1 in our context. Moreover, a gradient-based implementation of this approach can be found in [46]. Finally, we refer the reader to [47] for a review of several aspects of numerical approaches for mean-field control problems.

**Algorithm 1:** Mean-Field Equilibrium

**Initialization:**

1 Generate initial transmit power $P_0$ ;

**Learning pattern:**

2 Find $m$ using (22) with initial condition $m_0$ ;
3 Find $p_s$ and $\pi_a$ by solving algorithm 2 ;
4 Estimate interference $I_{mf}$ using (16);
5 Find $\mu$ using (26) with final condition (27);
6 Update transmit power $P$ using (25);
7 Repeat until convergence : go to step 2;

## 4. Grant-Free Markov Chain Model

In this section, we present a discrete-time Markov chain to derive a closed-form for the probability of having at least one packet in an IoT device's queue, which implies the probability of attempting uplink transmissions. To determine this probability analytically, we must first acquire the successful transmission probability $p_s$, which is defined as the probability that a packet is successfully transferred when an IoT device executes an uplink transmission.
It is worth mentioning that in this section, we will assume that IoT devices use their optimal power allocation strategy.
In our Markov chain model, each state represents the queue length of a generic IoT device, where the queue length implies the number of packets in the queue. The state space $S$ can be defined as

$$S = \{0, 1, 2, \dots, M\}, \tag{29}$$

where $M$ represents the queue size. Let's $\pi_i(n)$ be the state probability that the queue length equals to $i$ at time $n \in \{0, T_f, 2T_f, \dots\}$, thus the distribution of the state probability at time $n$, $\boldsymbol{\pi}(n)$, can be denoted as

$$\boldsymbol{\pi}(n) = \left[\pi_0(n), \pi_1(n), \dots, \pi_M(n)\right]. \tag{30}$$

The state transition probability from state $j$ to state $k$, denoted as $q_{j,k}$, is given in (31).

### 4.1. Packet Transmission Success Probability

The successful transmission probability at a generic BS is used to estimate the successful transmission probabilities of all IoT devices in the networks. This means that such probabilities are independent of location and uncorrelated across time frames. Exploiting this approximation and accounting for the mean-field interference, the successful transmission probability is characterized by the following proposition.

**Proposition 2.** *The successful transmission probability of a generic IoT device whose generic BS is at the origin is*

$$p_s = \frac{\pi\lambda_s}{T_f}\sum_{j=0}^{J}\mathbb{P}[N_a = j]\times \int_0^{T_f}\int_0^{E_{max}}\left[\int_0^{\infty} e^{-ar^{\frac{\alpha}{2}}} e^{-br} dr\right] m^*(t,e)\, de\, dt, \tag{32}$$

*where*

$$\begin{cases} a = \dfrac{\theta(\sigma_0 + I^*_{mf}(t))}{P^*(t,e)}, \\ b = \pi\lambda_s. \end{cases} \tag{33}$$

*Under urban areas where $\alpha \simeq 4$, we have*

$$p_s = \frac{\pi\lambda_s}{T_f}\sum_{j=0}^{J}\mathbb{P}[N_a = j]\int_0^{T_f}\int_0^{E_{max}} g(a,b) m^*\, de\, dt, \tag{34}$$

*where*

$$g(a,b) = \sqrt{\frac{\pi}{a}}\exp\left(\frac{b^2}{4a}\right)Q\left(\frac{b}{\sqrt{2a}}\right), \tag{35}$$

*with*

$$Q(x) = \frac{1}{\sqrt{2\pi}}\int_x^{\infty} e^{-\frac{u^2}{2}}\, du. \tag{36}$$

*The Q function is the tail distribution of the standard normal distribution.*

Proof. The proof is given in Appendix A.2.

### 4.2. Steady-State Analysis

The steady-state distribution $\boldsymbol{\pi}$ for the $M$ state Markov chain with transition matrix [K], given by using $q_{i,j}$ as the $i_{th}$ row and $j_{th}$ column element, is a row vector that satisfies

$$\boldsymbol{\pi} = \boldsymbol{\pi}[\mathrm{K}], \quad \text{where} \quad \langle \boldsymbol{\pi}, \mathbf{1}\rangle = 1 \quad \text{and} \quad \pi_i \geq 0. \tag{37}$$

Thus, the steady-state probabilities can be expressed as

$$\pi_0 = \pi_0 q_{0,0} + \pi_1 q_{1,0}, \tag{38}$$

$$\pi_i = \pi_{i-1}q_{i-1,i} + \pi_i q_{i,i} + \pi_{i+1}q_{i+1,i}, \quad i \in [1, M-1],$$

$$q_{j,k} = \begin{cases} 1-p_a, & j=k=0 \\ p_a, & j=0, \quad k=1 \\ (1-p_a)p_b + (1-p_a)(1-p_b)(1-p_s) + p_a(1-p_b)p_s, & 1 \le j \le M-1, \quad k=j \\ 1-(1-p_a)(1-p_b)p_s, & j=k=M \\ p_a(1-p_b)(1-p_s) + p_a p_b & 1 \le j \le M-1, \quad k=j+1 \\ (1-p_a)(1-p_b)p_s, & 1 \le j \le M, \quad k=j-1 \\ 0, & \text{Otherwise.} \end{cases} \tag{31}$$

(39)

and

$$\pi_M = \pi_{M-1} q_{M-1,M} + \pi_M q_{M,M}. \tag{40}$$

Thus with above equations, we have

$$\pi_i = \begin{cases} \pi_0, & i = 0, \\ \pi_0 \prod_{j=0}^{i-1} \left( \frac{q_{j,j+1}}{q_{j+1,j}} \right), & i \in [1, M]. \end{cases} \tag{41}$$

By using the normalization condition $\sum_i \pi_i = 1$, $\pi_0$ can be expressed as

$$\pi_0 = \left( 1 + \sum_{i=1}^{M} \prod_{j=0}^{i-1} bigg( \frac{q_{j,j+1}}{q_{j+1,j}} \right) \Bigg)^{-1}. \tag{42}$$

Therefore, the probability that an IoT device is active, i.e., there exist at least a packet in its queue, is given by

$$\pi_a = 1 - \pi_0 = 1 - \left( 1 + \sum_{i=1}^{M} \prod_{j=0}^{i-1} \left( \frac{q_{j,j+1}}{q_{j+1,j}} \right) \right)^{-1}. \tag{43}$$

Note that because the equations (32) and (43) form a nonlinear system given $P^*$ and $m^*$, which means that there is no closed-form solution that can be directly calculated. Therefore, $\pi_a$ and $p_s$ must be computed numerically. The algorithm 2 summarizes the numerical approach. The convergence of the algorithm 2 is guaranteed due to the uniqueness of steady-state probabilities. In other words, given a fixed set of system parameters, the steady-state probabilities of the Markov chain are unique and can be obtained through numerical methods. Our iterative approach is based on updating the transmission success probability and the probability that an IoT device is active in each iteration until convergence is achieved.

**Algorithm 2:** Learning $p_s$ and $\pi_a$

**Initialization:**
1. Generate initial success probability $p_s$ ;
2. Calculate $\pi_a$ using (43) ;

**Learning pattern:**

3. Update $p_s$ using (32);
4. Update $\pi_a$ using (43) ;
5. Repeat until convergence : go to step 3;

### 4.3. Performance Metrics

By solving algorithm 1, we will obtain the optimal power allocation strategy as well as the Markov chain steady-state probabilities. As a result, we can now exploit them to estimate the average steady-state performance of grant-free access in terms of throughput, queue size, number of transmissions, and delay.
Note that, because of the Bernoulli arrival process, the packet generation is a geometric inter-arrival process with parameter $p_a$, i.e. the interval between two consecutive packet arrivals is a geometric random variable. As a result, Geo/Geo/1/M queue can be used to represent the discrete-time queuing system of a generic IoT device. More precisely, since the success probability $p_s$ is uncorrelated across time frames, the service time (in number of frames) is also a geometric random variable with the parameter $(1 - p_b)p_s$.
- **Average throughput rate (service rate):** The average throughput rate, denoted as $T_h$, experienced by a generic IoT device writes

$$T_h = (1 - p_b)p_s. \tag{44}$$

It should be noted that access barring can affect the average throughput rate. More precisely, for a given $p_b$, the system can operate in either a saturated or unsaturated state. When $p_a > (1 - p_b)p_s$, the system is saturated, which means that all IoT devices have at least one packet to transmit. In contrast, when $p_a < (1 - p_b)p_s$, the system works in an unsaturated state.

- **Average number of transmissions (service time):** Let $N_t$ represent the number of transmission attempts before a generic IoT device's packet is successfully transmitted, which follows a geometric distribution with parameter $T_h$, and its probability mass function can be written, for $k \in \{1, 2, \dots\}$, as

$$f_{N_t}(k) = \left(1 - T_h\right)^{k-1} T_h. \tag{45}$$

Thus, the average number of transmissions per packet is given by:

$$\mathbb{E}[N_t] = \frac{1}{T_h}. \tag{46}$$

- **Average queue size:** At steady state, the average queue size of a generic IoT device can be estimated by

$$Q = \sum_{k=1}^{M} k\pi_k. \tag{47}$$

- **Average delay:** Let $D$ denote the average delay experienced by a given packet at a generic IoT device, i.e., the average number of frames that a packet spent in the queue until successful transmission, which may be represented as

$$D = \underbrace{Q\mathbb{E}[N_t]}_{\text{queuing delay}} + \underbrace{\mathbb{E}[N_t]}_{\text{transmission delay}}. \tag{48}$$

It should be emphasized that when the system is saturated, the average delay increases dramatically.

## 5. Numerical Analysis

In this section, we offer some explanations below on how to numerically solve the algorithm 1 using a finite difference method and we present numerical results.

### 5.1. Finite Difference Method

In order to numerically solve our algorithm, we consider a discretized grid within $[0, T_f] \times [0, E_{max}]$. Let us consider two positive integers, $X$ and $Y$. We define the time and space steps by $\delta t = T_f/X$, $\delta E = E_{max}/Y$, and for $n = 0, \dots, X$, $i = 0, \dots, Y$, we denote $f_i^n$ the numerical approximations of $f(n\delta t, i\delta E)$ for any function $f$. The FPK equation (22) is computed iteratively using a upwind-type finite difference scheme by:

$$m_i^{n+1} = m_i^n + \frac{\delta t}{\delta E}\left(P_{i+1}^n m_{i+1}^n - P_i^n m_i^n\right). \tag{49}$$

**Table 1**
Parameters for Numerical Results

| Parameter | Values | Description |
|---|---|---|
| $\lambda_s$ | 1,5,10,20 BS/k$m^2$ | BS density |
| $\lambda_u$ | 3000 IoT/k$m^2$ | IoT density |
| $p_b$ | 0.1 | Barring probability |
| $p_a$ | 1-60 packet/hour | Arrival rate |
| $\theta$ | 10 | SINR threshold |
| $J$ | 1,3,5,7 | MPR capability |
| $L$ | 30 | Number of channels |
| $M$ | 10 | The queue size |
| $P_{max}$ | 0.025 W | Maximum power |
| $T_f$ | 10 ms | Frame duration |
| $E_{max}$ | 0.1 mJ | Maximum energy |
| $\sigma_0$ | $10^{-23}$ W | Noise power |
| $\alpha$ | 4 | Path-loss exponent |

Moreover, for an arbitrary point $(n, i)$, the optimality conditions (25), (26) are discretized as follows:

$$\frac{\partial F_i^n}{\partial P_i^n} - \frac{\left(\mu_i^n - \mu_{i-1}^n\right)}{\delta E} = 0, \tag{50}$$

$$\mu_i^{n-1} = \mu_i^n - \frac{\delta t}{\delta E} P_i^n \left(\mu_i^n - \mu_{i-1}^n\right) + F_i^n \delta t, \tag{51}$$

where

$$F_i^n = (1-p_s)P_i^n - p_s \log_2\left(1 + \frac{P_i^n(2\sqrt{\lambda_s})^\alpha}{\sigma_0 + I_{mf}(t)}\right). \tag{52}$$

### 5.2. Numerical Results

We present numerical results on the performance of the algorithm 1. For all simulations, we choose $X = 100$ and $Y = 30$ to form a discretized space $X \times Y$, and we consider a uniform initial energy distribution $m_0$. Table 1 shows the typical values of parameters used for numerical results.

The Figure 1 shows the optimal power allocation strategy and the successful transmission probability for different BS densities. In this simulation, we set the arrival rate $p_a$ to 1 packet / 5 min and the MPR capability $J$ to 3. The IoT devices can adjust their transmit power based on their available energy at any given time. As the network becomes denser, both the successful

**Figure 1:** Optimal Power Allocation Strategies and successful transmission probability for different BS densities with $p_a = 1$ packet / 5 min and $J = 3$.

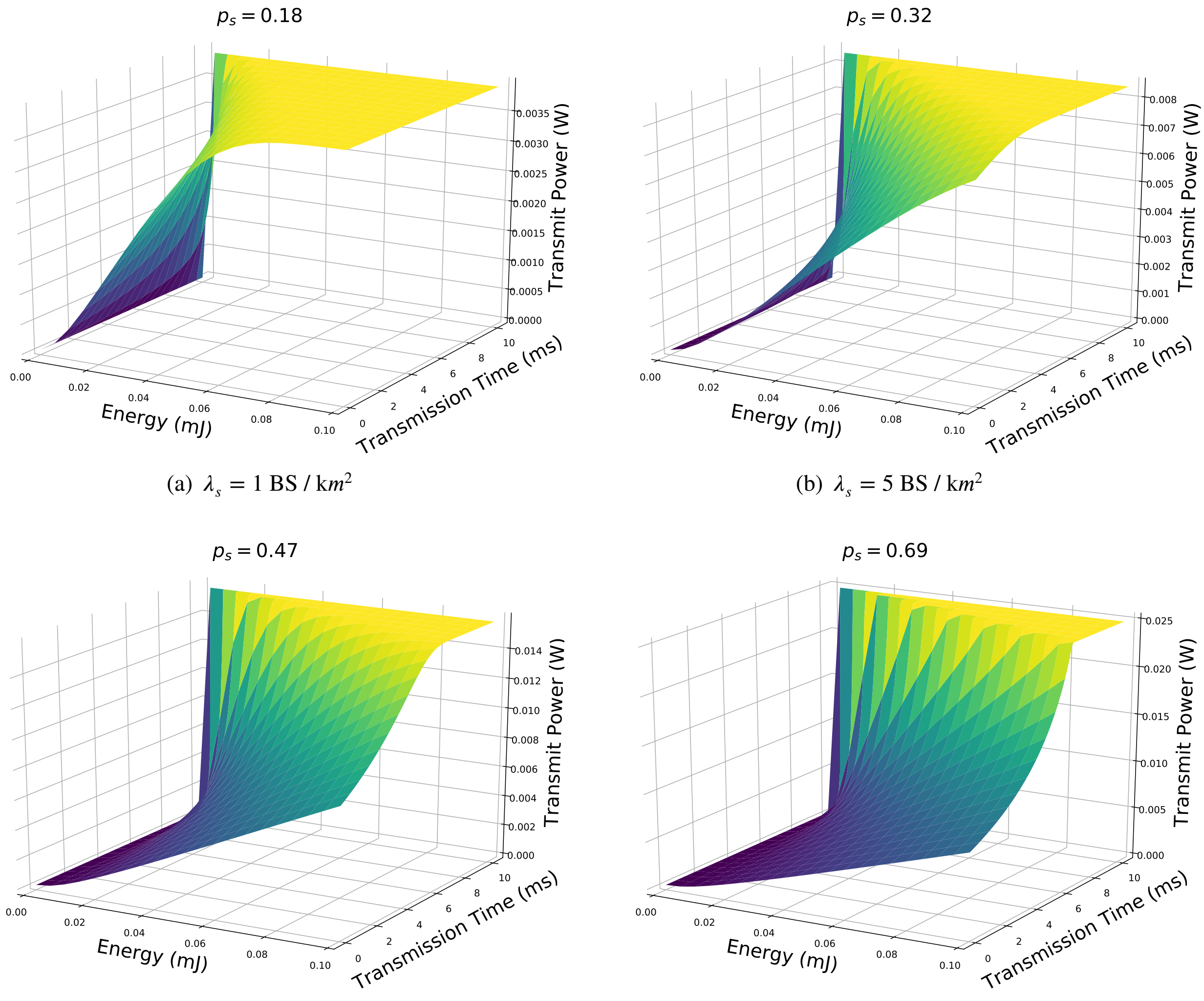

(a) $\lambda_s = 1$ BS / k$m^2$

(b) $\lambda_s = 5$ BS / k$m^2$

(c) $\lambda_s = 10$ BS / $km^2$

(d) $\lambda_s = 20$ BS / $km^2$

transmission probability and the number of points for which $P^* = P_{max}$ increase. It is worth noting that at the end of transmission, IoT devices with lower energy budgets empty their batteries, lowering the average network interference. As a result, IoT devices find that they can boost their transmit power to maximize their throughput.

The mean-field at the equilibrium is shown in Figure 2 for two BS densities, and Figure 3 illustrates a cross-section of this mean-field at fixed energy levels. Both figures illustrate the evolution of the energy distribution with time. As stated before, a uniform initial energy distribution $m^0$ is considered, i.e., the initial probabilities are similar for all energy budgets. It can be seen that the fraction of IoT devices with higher energy reduces over time, especially in the case of a dense network ($\lambda_s = 20$ BS / k$m^2$). This is because IoT devices increase their transmit power since they can achieve a better success probability. Meanwhile, when $\lambda_s$ is set to 1 BS/k$m^2$, approximately 13% of the IoT devices use up their whole energy budget during transmission. In contrast, if the IoT devices transmit at maximum power $P_{max}$, they will all consume their entire energy budget before the end of the considered time frame $T_f$.

The benefits of increasing base station density and improving multi-packet reception capability are highlighted in Figure 4 and Figure 5 by investigating the average delay. More precisely, Figure 4 illustrates the average delay in number of frames as a function of BS densities for various arrival rates. The simulation sets the MPR capability $J$ to 3, and it shows that as the BS density increases, the average delay reduces. This is due to the fact that a higher density of BSs shortens the distance between an IoT device and its serving BS, resulting in more reliable transmission. The figure further demonstrates that as the BS density grows, the difference in the average delay between different arrival rate scenarios narrows. This is because a higher density of BSs can handle more traffic and reduces network congestion. Hence, the delay experienced by an IoT device is reduced, regardless of the arrival rate. Moreover, Figure 5 shows the average delay as a function of BS densities for various multi-packet reception capabilities. In this simulation, we set the arrival rate $p_a$ to 1 packet / min. It shows that as the MPR capability improves, the average delay reduces. This is because the MPR capability allows the base station to receive multiple packets simultaneously, increasing the throughput and reducing the average delay. However, the figure clearly demonstrates that the advantage of MPR capability fades as the network becomes denser. This is due to the fact that the number of resources such as channel and MPR increases with the number of BSs. Hence, the additional resources have little or no effect on reducing the average delay.

**Figure 2:** Mean-field at the equilibrium for different BS densities with $p_a = 1$ packet / 5 min and $J = 3$.

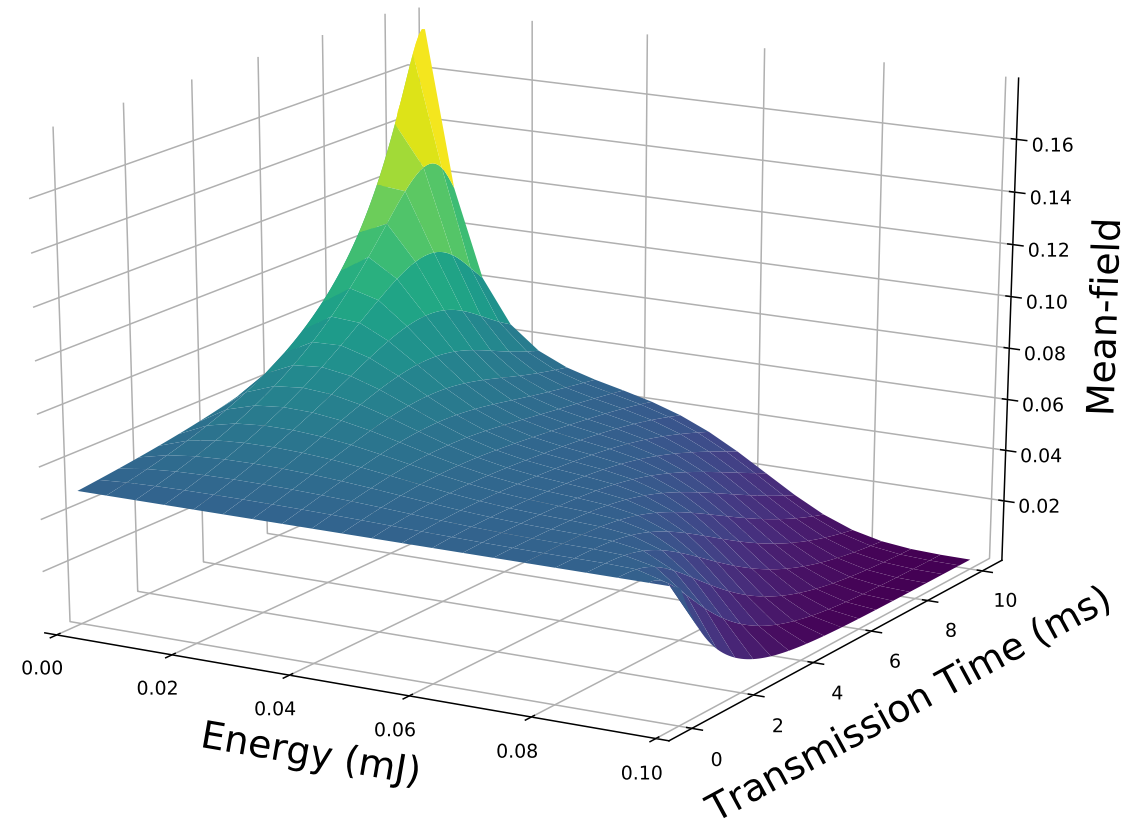

(a) $\lambda_s = 1$ BS / km$^2$

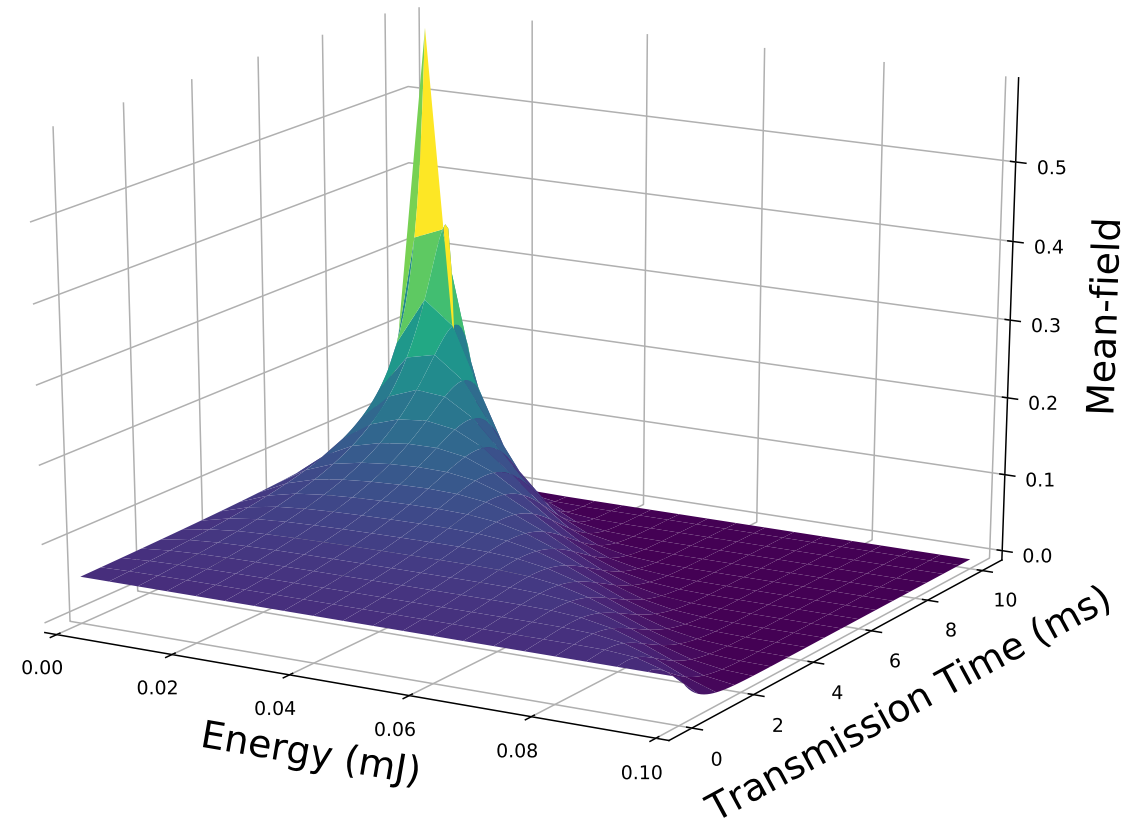

(b) $\lambda_s = 20$ BS / km$^2$

**Figure 3:** Cross-section of the mean-field at the equilibrium for different BS densities and energy levels with $p_a = 1$ packet / 5 min and $J = 3$.

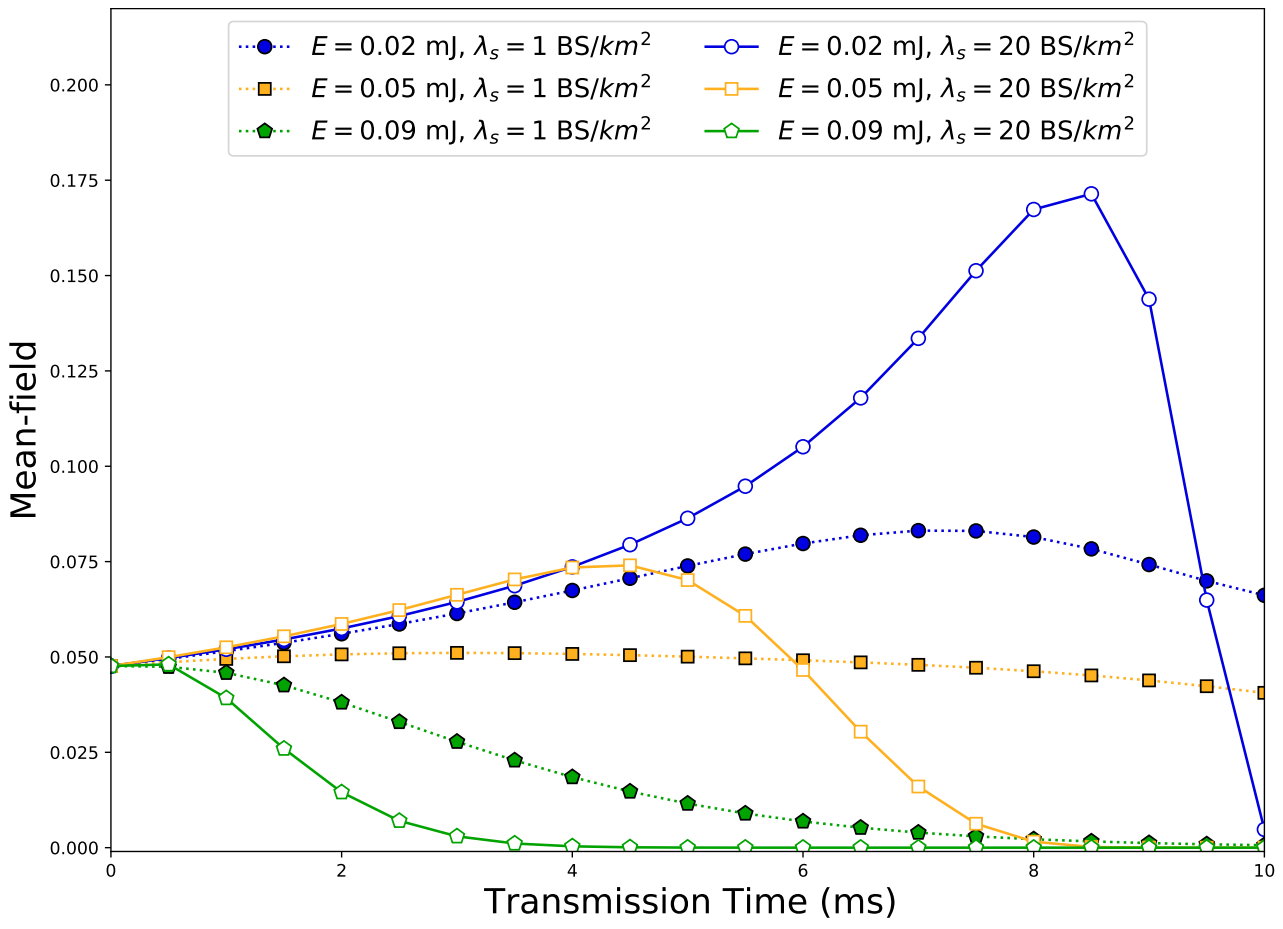

**Figure 4:** Average delay as a function of BS densities for various arrival rates with $J = 3$.

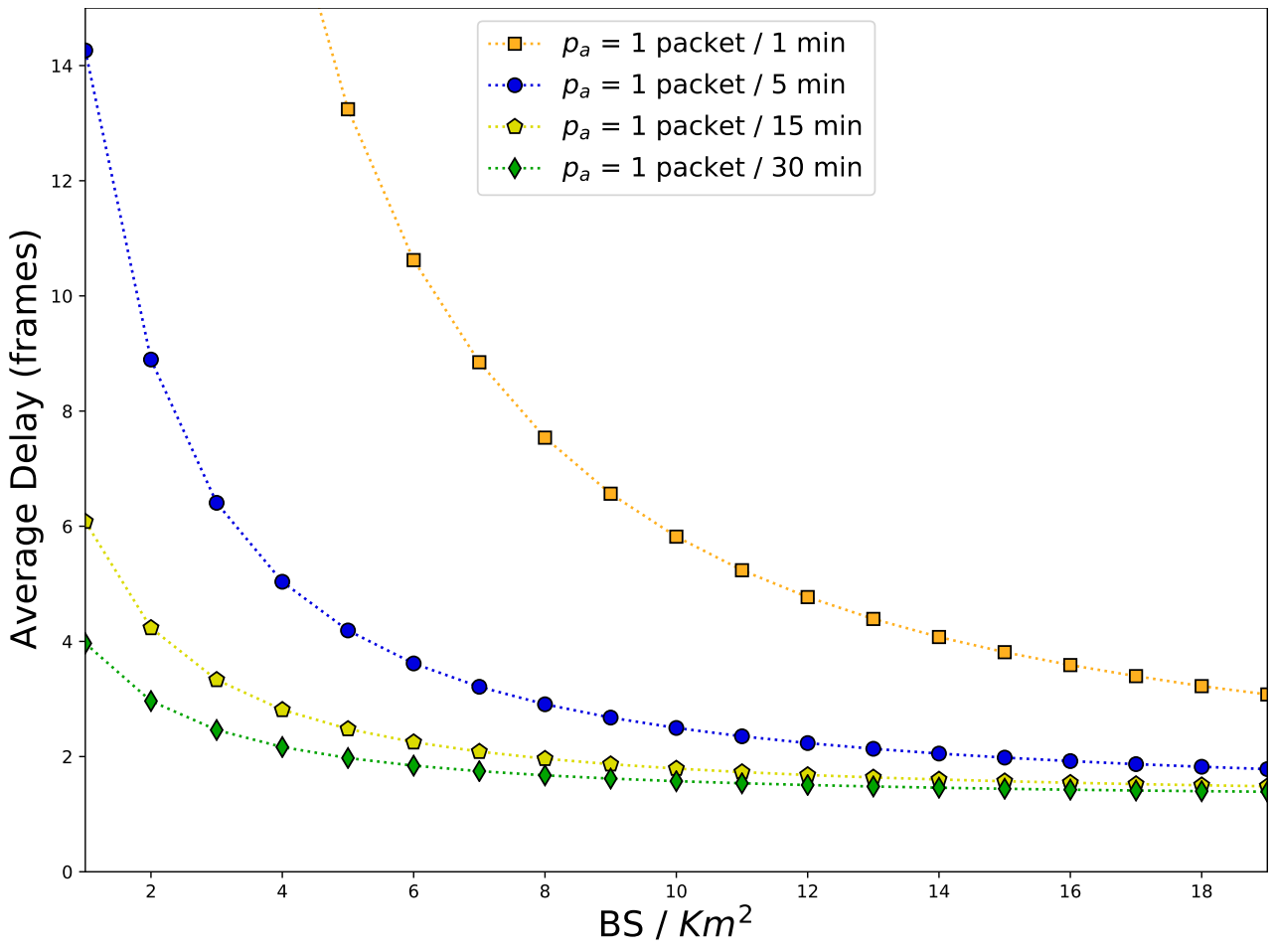

The influence of IoT density on average delay is highlighted in Figure 6 and Figure 7. In Figure 6, the average delay is illustrated as a function of IoT density for different arrival rates, while the BS density is set to 10 BS/km$^2$, and the MPR capability $J$ is set to 3. The figure indicates that as the IoT density increases, the average delay also increases. This is because an increase in the IoT density leads to more IoT devices contending for the same resources, resulting in congestion and increased delay. However, in low-traffic conditions, the average delay remains acceptable even as the IoT density increases. The figure suggests that the arrival rate has a more significant impact on the average delay than the IoT density. The Figure 7, on the other hand, shows the average delay as a function of IoT density for various BS densities, with the arrival rate fixed at 1 packet/min. The figure illustrates that as the IoT density increases, the average delay also increases, regardless of the BS density.

**Figure 5:** Average delay as a function of BS densities for various multi-packet reception capabilities with $p_a = 1$ packet / 1 min.

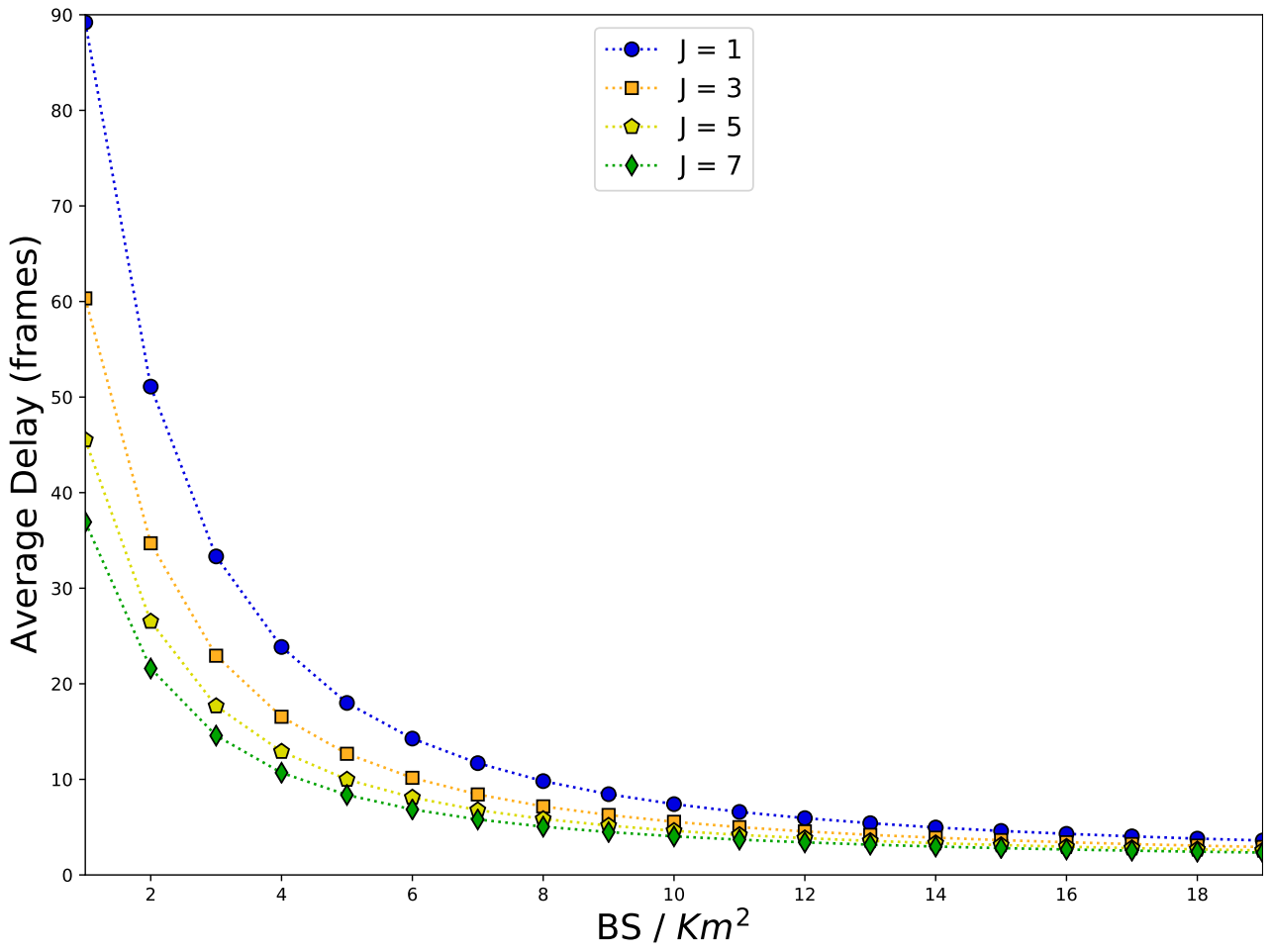

Finally, the influence of IoT density and arrival rate on the success transmission probability is highlighted in Figure 8 and Figure 9. It's worth noting that the success probability in these figures represents the probability of successfully transmitting a packet in a single frame. This probability can be compared to the success probability of the slotted Aloha protocol, which is approximately 18% for a single device. The Figure 8 shows the success probability as a function of the arrival rates for different MPR capabilities and BS densities. The figure indicates that in low-traffic conditions, a high packet transmission success probability can be maintained without the need for network densification or MPR capability. However, under high-traffic conditions, network densification and/or MPR capability can help maintain a high success probability. On the other hand, Figure 9 illustrates the success probability as a function of IoT densities for different arrival rates and BS densities, with the MPR capability set to 3. The figure shows that as the IoT density increases, the packet transmission success probability decreases, which is due to increased contention for network resources. The figure also suggests that network densification can help improve the success probability, but the improvement diminishes as the IoT density increases.

**Figure 6:** Average delay as a function of IoT device densities for various arrival rates with $\lambda_s = 10$ BS / km$^2$ and $J = 3$.

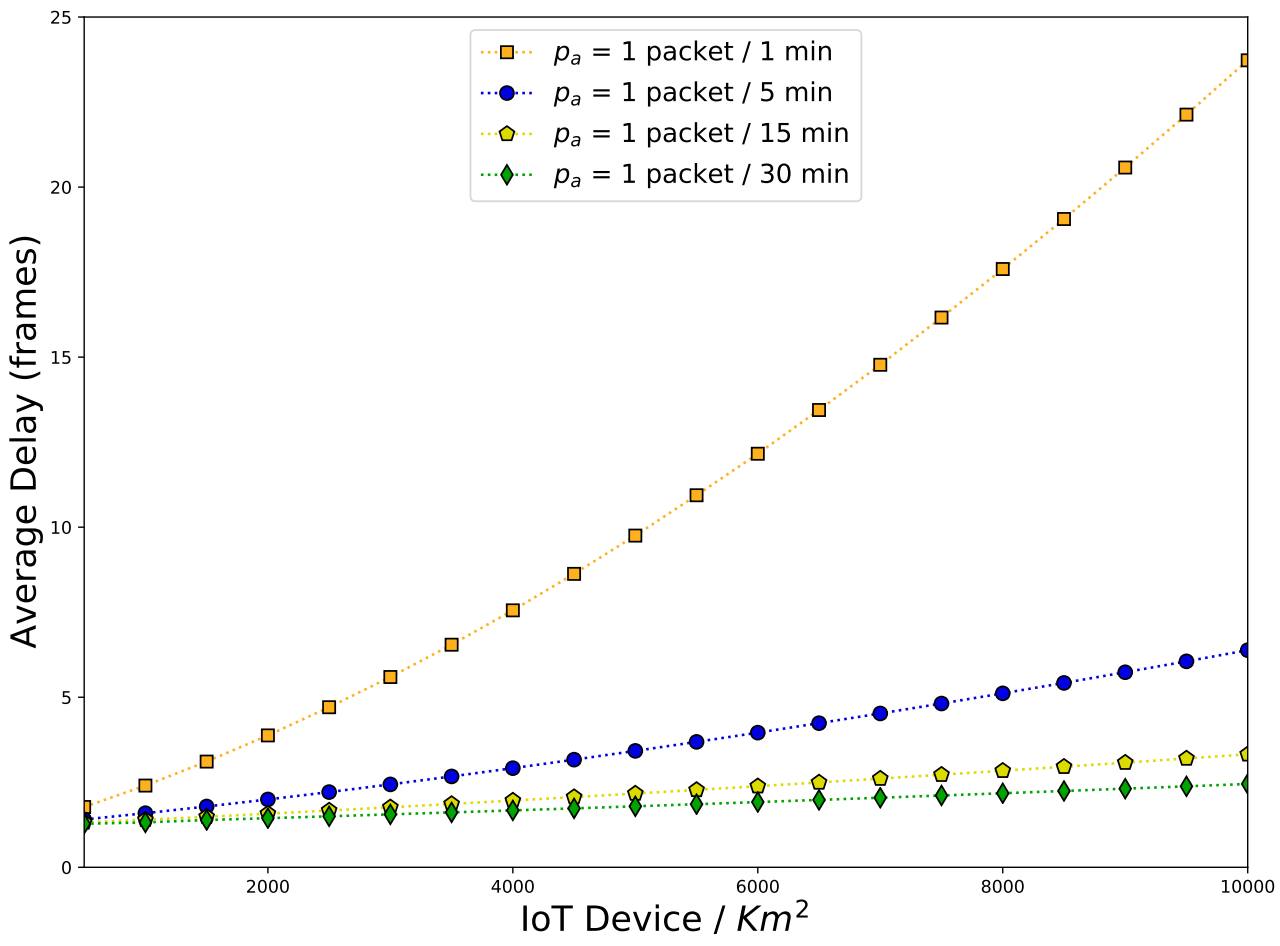

**Figure 7:** Average delay as a function of IoT device densities for various BS densities with $p_a = 1$ packet / 1 min and $J = 3$.

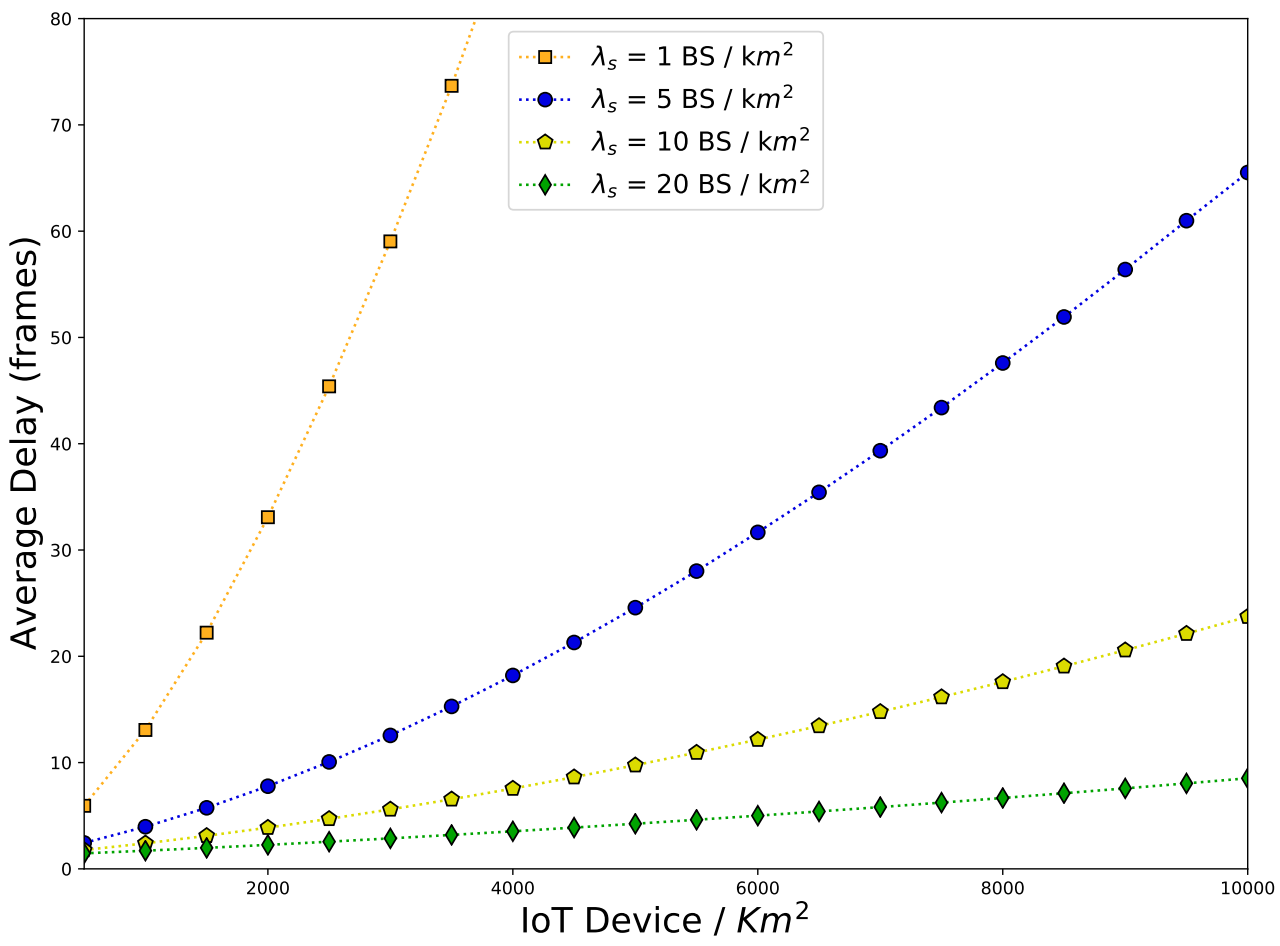

## 6. Conclusion

The paper investigates the grant-free access with multi-packet reception capabilities for ultra-low-end IoT applications with a focus on energy usage limits and small data sizes. The paper presents a distributed power allocation strategy for a large-scale massive IoT environment to meet the throughput expectations of IoT devices while minimizing their energy usage. The mean-field framework is used to capture population behavior, and the Markov chain framework is used to derive the successful transmission probability in a massive grant-free IoT network. The success probability is influenced by multiple factors, such as IoT density, arrival rate, base station density, and multi-packet reception capability. The numerical findings show the optimal power allocation strategy under various BS density conditions, as well as the benefits of network densification and multi-packet reception capability by investigating average delay and packet transmission success probability. In particular, the simulation highlights the fact that increasing the base station density and improving the multi-packet reception capability can reduce the average delay, but their effectiveness may depend on the network density and available resources. It also provide insights into the impact of IoT density on the performance of IoT networks and highlight the importance of optimizing network resources to minimize the average delay.

**Figure 8:** Success probability as a function of arrival rates for various multi-packet reception capabilities and BS densities.

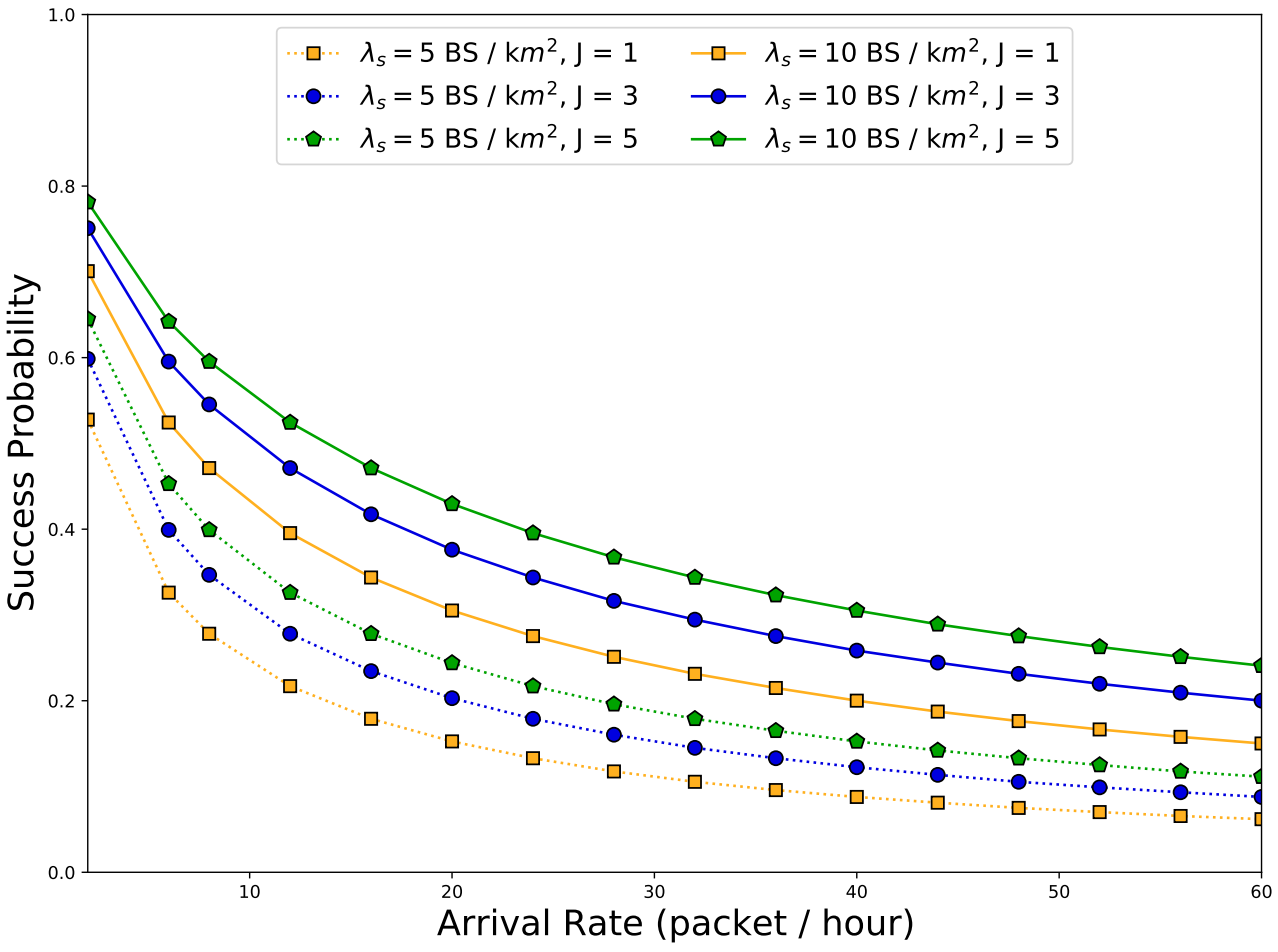

**Figure 9:** Success probability as a function of IoT device densities for various arrival rates and BS densities with $J = 3$.

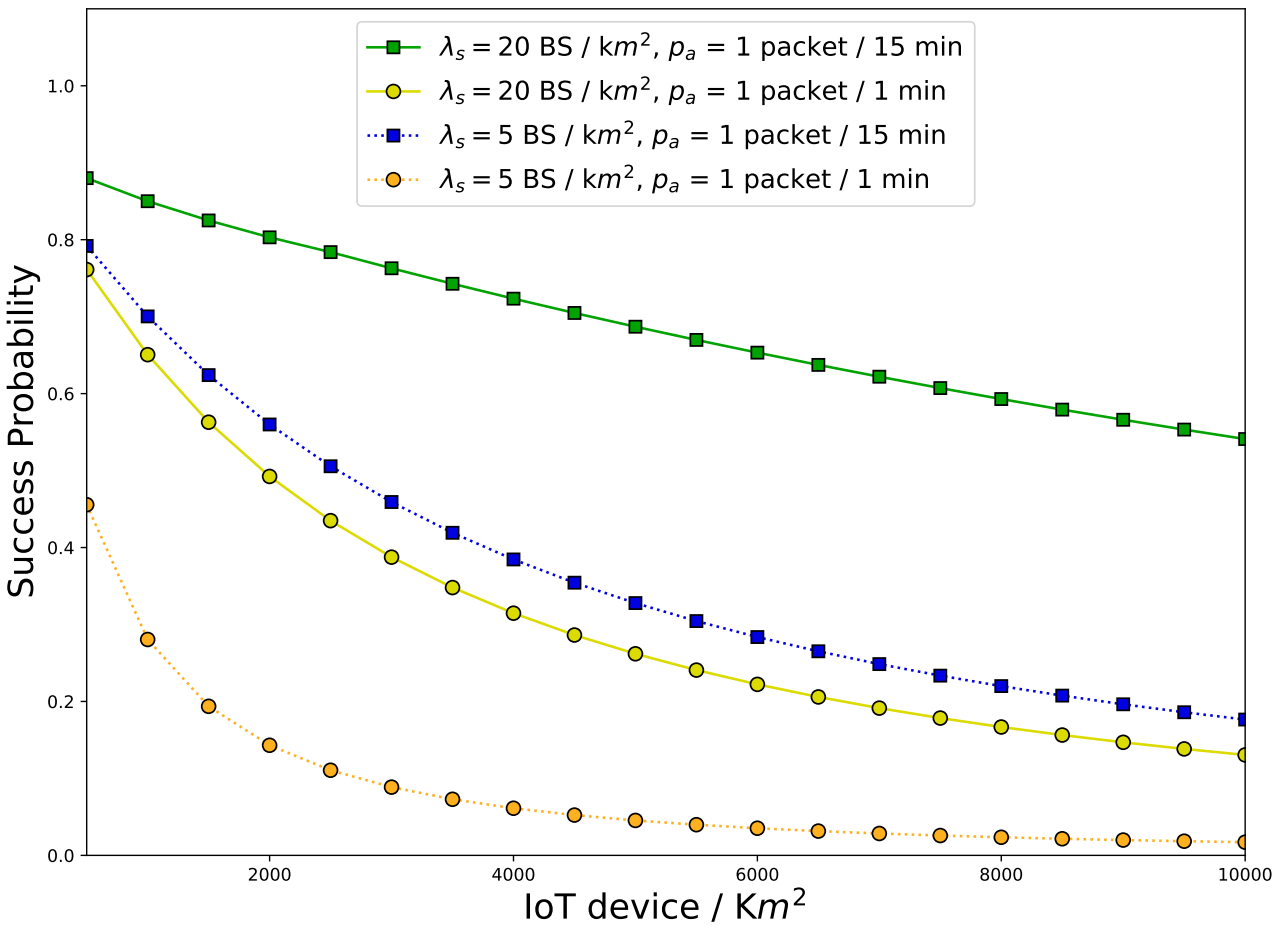

# A. Appendix

## A.1. Proof of proposition 1

Without loss of generality, we derive the finite mean-field interference for an IoT device $i$ whose BS is at the origin.
The mean-field interference in $C_R$ for a given time instant $t \in [0, T_f]$ is expressed as:

$$I_{i,mf}(t) = \mathbb{E}\left[\sum_{j=1, j\neq i}^{|\mathcal{N}_R|} P(t, E_j(t)) H_{i,j}(t) D_{i,j}(r)\right]. \quad (53)$$

It is worth noting that, in the mean-field regime, the IoT devices become essentially indistinguishable, and a single IoT device has a negligible effect on the overall mass behavior. Thus, we can focus on a generic IoT device while dropping the index $i$ from (53). Since the transmit power of a generic IoT device is independent of the point process, and $h$ is exponentially distributed with unity mean, then the previous formula writes

$$I_{mf}(t) = \mathbb{E}\left[P(t, E)\right] \mathbb{E}_{\phi_a}\left[\sum_{j=1, j\neq i}^{|\mathcal{N}_R|} D(r)\right], \quad (54)$$

where

$$\mathbb{E}[P(t, E)] = \int_0^{E_{max}} P(t, e) m(t, e)\, de. \quad (55)$$

Then, by using Campbell's formula, we write:

$$\mathbb{E}_{\phi_a}\left[\sum_{j=1, j\neq i}^{|\mathcal{N}_R|} r^{-\alpha}\right] = 2\pi\lambda_u \frac{(1 - p_b)\pi_a}{L} \int_0^R D(r) r dr. \quad (56)$$

Since the received power cannot be larger than transmit power, the path-loss is assumed to be 1 when $r < 1$. Then, we have

$$\int_0^R D(r) r dr = \int_0^1 r dr + \int_1^R r^{1-\alpha} dr = \frac{1}{2} + \frac{1 - R^{\alpha-2}}{2 - \alpha}. \quad (57)$$

Finally, by considering a large scale network and taking $R \to \infty$ concludes the proof.

### A.2. Proof of proposition 2

The successful transmission probability is given by

$$p_s = \underbrace{\sum_{j=0}^{J} \mathbb{P}[N_a = j]}_{\text{probability of no packet collision}} \times p_\theta \quad (58)$$

where $p_\theta$ is the probability that the SINR of a generic IoT device, whose generic BS is at the origin , is greater than $\theta$ over a given frame with duration $T_f$ expressed as

$$\begin{aligned} p_\theta &= \frac{1}{T_f}\int_0^{T_f} \mathbb{P}\left[\frac{P^*(t,e)Hr^{-\alpha}}{\sigma_0 + I} \geq \theta\right] dt \\ &\approx \frac{1}{T_f}\int_0^{T_f} \mathbb{P}\left[\frac{P^*(t,e)H}{r}^{-\alpha} \sigma_0 + I^*_{mf} \geq \theta\right] dt. \end{aligned} \quad (59)$$

Conditioning on the energy and the distance from a generic IoT device to its nearest BS, we get

$$\begin{aligned} p_\theta &= \frac{1}{T_f}\int_0^{T_f} \mathbb{E}\left[\mathbb{P}\left[\frac{P^*(t,e)Hr^{-\alpha}}{\sigma_0 + I^*_{mf}} \geq \theta \middle| r,e\right]\right] dt \\ &= \frac{1}{T_f}\int_0^{T_f}\int_0^{E_{max}}\left[\int_0^{\infty} \mathbb{P}\left[H \geq ar^\alpha\right] f_r dr\right] m^*\, de\, dt, \end{aligned} \quad (60)$$

where $a = \theta(\sigma_0 + I^*_{mf}(t))/P^*(t,e)$.
Using the fact that $H \sim \exp(1)$, we have

$$\mathbb{P}\left[H \geq ar^\alpha\right] = e^{-ar^\alpha}. \quad (61)$$

Thus, replacing the probability density function $f_r$ by its expression given in equation (2), yields

$$p_\theta = \frac{2\pi\lambda_s}{T_f}\int_0^{T_f}\int_0^{E_{max}}\left[\int_0^{\infty} e^{-ar^\alpha} e^{-br^2} r dr\right] m^*(t,e)\, de\, dt. \quad (62)$$

where $b = \pi\lambda_s$.
Using the substitution $s = r^2$ in the inside integral of (62) and combining with (58), we obtain the result.
Finally, in the special case where $\alpha = 4$, we have the following result

$$\int_0^{\infty} e^{-ar^2} e^{-br} dr = \sqrt{\frac{\pi}{a}} \exp\left(\frac{b^2}{4a}\right) Q\left(\frac{b}{\sqrt{2a}}\right), \quad (63)$$

where

$$Q(x) = \frac{1}{\sqrt{2\pi}} \int_x^{\infty} e^{-\frac{u^2}{2}} du. \quad (64)$$

This concludes the proof.

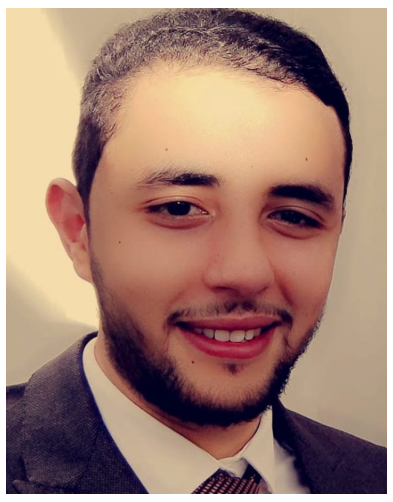

**Sami Nadif** (Student Member, IEEE) obtained his Master degree in applied mathematics and computation in 2017, from the Faculty of Science and Technology, Hassan First University of Settat, Morocco. Sami holds a part-time lecturer position at the Moroccan School of Engineering Sciences since 2020. He is currently pursuing his Ph.D. in the Department of Computing, Networks Mobility and Modeling at Hassan First University. His main research interests include self-organizing networks, ubiquitous networks, the internet of things, game theory, and learning approaches for ultra-dense wireless networks.

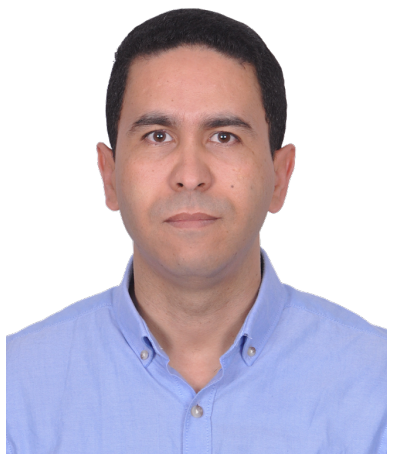

**Essaid Sabir** (Senior Member, IEEE) received the B.Sc. degree and the M.Sc. degree in ECE from Mohammed V University of Rabat in 2004 and 2007 respectively, and the Ph.D. degree (Hons.) in networking and computer engineering from Avignon University, France, in 2010. He held a non-tenure-track Assistant Professor position at Avignon University, from 2009 to 2012. He has been a Professor at ENSEM - Hassan II university of Casablanca, until late 2022, where he was leading the NEST research Group. He is currently a substitute professor with the department of computer science, Université du Québec à Montréal. He is/was the main investigator of many research projects, and has been involved in several other (inter)national projects. His research interests include 5G/6G, Cloud/Fog/Edge, IoT, URLLC, AI/ML, and networking games. His work has been awarded in many international venues (5GWF'21, WF-IoT'20, IWCMC'19, etc.). To bridge the gap between academia and industry, he founded the International Conference on Ubiquitous Networking (UNet) and co-founded the WINCOM conference series. He serves as an associate/guest editor for many journals. He organized numerous events and played executive roles for many major venues.

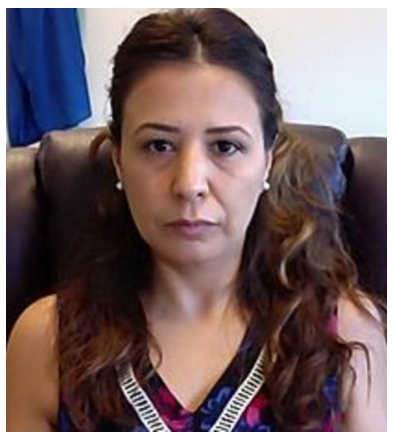

**Halima Elbiaze** (Senior Member, IEEE) received the B.Sc. degree in applied mathematics from Mohammed V University, Rabat, Morocco, in 1996, the M.Sc. degree in telecommunication systems from the Université de Versailles, Versailles, France, in 1998, and the Ph.D. degree in computer science from the Institut National des Télécommunications, Paris, France, in 2002. Since 2003, she has been with the Department of Computer Science, Université du Québec à Montréal, Montreal, QC, Canada, where she is currently an Associate Professor. She has authored or co-authored many journal and conference papers. Her current research interests include network performance evaluation, traffic engineering, and quality of service management in optical and wireless networks.

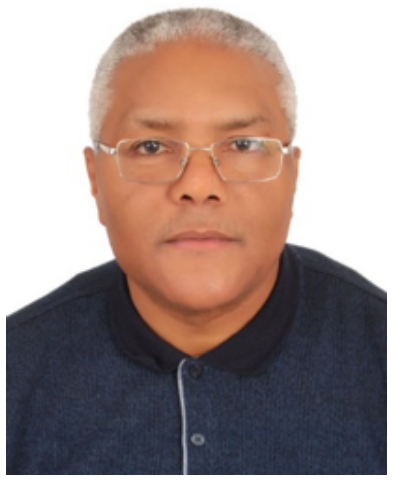

**Abdelkrim Haqiq** (Senior Member, IEEE) has a High Study Degree (Diplôme des Etudes Supérieures de troisième cycle) and a PhD (Docrotat d'état), both in modeling and performance evaluation of computer networks communications, from Mohammed V University of Rabat, Morocco. Since September 1995 he has been working as a Professor at the department of Applied Mathematics and Computer at the Faculty of Sciences and Techniques, Settat, Morocco. He is the Director of Computer, Networks, Mobility and Modeling laboratory (IR2M). He is also a member of Machine Intelligence Research Labs (MIR Labs), Washington-USA, and a member of the International Association of Engineers. He was a co-director of a NATO Multi-Year project entitled "Cyber-Security Analysis and Assurance using Cloud-Based Security Measurement system" (SPS-984425). His interests lie in the areas of modeling and performance evaluation of communication networks, mobile communications networks, cloud computing and security, emergent technologies, Markov Chains and queueing theory, Markov decision processes theory, and game theory. He has (co)authored more than 170 journal/conference papers. He serves as associate editor, editorial board member, international advisory board member, and editorial review board member of many international journals. He was a chair and a technical program committee chair/member of many international conferences. He was also a Guest Editor for many journals, books and conference proceedings.